\documentclass[conference]{IEEEtran}

\usepackage{amssymb,amsmath}
\usepackage{hyperref}
\usepackage{caption}
\usepackage{amsfonts}
\usepackage{multirow}
\usepackage{color}
\usepackage[tableposition=top]{caption}
\ifCLASSINFOpdf
   \usepackage[pdftex]{graphicx}
\else
\fi
\newcommand{\ie}{i.e.}
\newcommand{\eg}{e.g.}
\newcommand{\mbb}[1]{\mathbb{#1}}
\newcommand{\mc}[1]{\mathcal{#1}}

\newcommand{\SSIM}{\mathrm{SSIM}}
\newcommand{\SNR}{\mathrm{SNR}}

\begin{document}

\title{A Variational Auto-Encoder Approach for Image Transmission in Wireless Channel}

\author{
\IEEEauthorblockN{Amir Hossein Estiri, Mohammad Reza Sabramooz, Ali Banaei, Amir Hossein Dehghan, Benyamin Jamialahmadi\\ Mahdi Jafari Siavoshani}
\IEEEauthorblockA{Information, Network, and Learning Lab, Sharif University of Technology, Tehran, Iran
}
}

\maketitle

\begin{abstract}
Recent advancements in information technology and the widespread use of the Internet have led to easier access to data worldwide. As a result, transmitting data through noisy channels is inevitable. Reducing the size of data and protecting it during transmission from corruption due to channel noises are two classical problems in communication and information theory. Recently, inspired by deep neural networks' success in different tasks, many works have been done to address these two problems using deep learning techniques.

In this paper, we investigate the performance of variational auto-encoders and compare the results with standard auto-encoders. Our findings suggest that variational auto-encoders are more robust to channel degradation than auto-encoders. Furthermore, we have tried to excel in the human perceptual quality of reconstructed images by using perception-based error metrics as our network's loss function. To this end, we use the structural similarity index (SSIM) as a perception-based metric to optimize the proposed neural network. Our experiments demonstrate that the SSIM metric visually improves the quality of the reconstructed images at the receiver.
\end{abstract}


\IEEEpeerreviewmaketitle

\section{Introduction}
It has been a long time since Shannon introduced a mathematical model for communication transmission and investigated the point-to-point channel in his seminal paper \cite{shannon1948} in 1948. There have been many investigations and extensions during these years that considered various types of channels by applying tools from information theory, estimation theory, and statistics that have led to successful practical designs (\eg, see \cite{EIT_CoverBook,NIT_ElGamal_Kim}, and the references therein).

Although the fields of information and communication theory have flourished a lot, the optimal solution for many of the transmission scenarios, even those that are not very complicated, remains unknown for many years. This makes the goal of finding optimal (or even practical) solutions to more general problems in network information theory out of reach in the near future.

Recently, inspired by the success of applying deep learning (DL) techniques in various research fields, it has been suggested to design an end-to-end communication system by modeling it through a deep neural network and training this model for many iterations over the channel realizations (\eg, see \cite{OShea_TCCN17,DeepJSCC_Gunduz_TCCN19,simeone2018survey,wang2017DL4WPL,mao2018survey}, and the references therein). Using such an approach, even if there exists no mathematical model for a channel, or if the model is analytically complicated to deal with, as far as we can simulate that channel or have access to measured noise data, we can employ DL techniques to provide a practical solution that works well in practice.

Many works have considered applying machine learning techniques to design various parts of a communication system, including equalization, quantization, decoding, demodulation, to name just a few (\eg, see \cite{ibnkahla2000applications,bkassiny2012survey}, and the references therein). However, to the best of our knowledge, \cite{OShea_TCCN17} was the first work that employed DL tools to design an end-to-end system by interpreting a communication system as an auto-encoder (AE) \cite{ballard1987modular,Goodfellow_DeepLearning_2016}. In this approach, various modules at the transmitter, including source coding, channel coding, and modulator, are modeled by a neural network called \emph{encoder} in AE. Similarly, the modules applied at the decoder, such as the demodulator, channel decoder, and decompressor, are also modeled by another neural network, called decoder in AE. Furthermore, the noisy channel can be considered as a non-trainable layer, and then the whole system is trained using the \emph{backpropagation} algorithm. It is shown in \cite{OShea_TCCN17} that some of the state-of-the-art results in the field of communication, in particular the results for the point-to-point channel, can be recovered using such a DL approach.

As a follow up to the above work, \cite{DeepJSCC_Gunduz_TCCN19} suggested a joint source-channel coding for wireless image transmission. This work also employs an auto-encoder structure that simultaneously learns how to compress the images and map the compressed symbols to channel signals. The results show that the proposed scheme in \cite{DeepJSCC_Gunduz_TCCN19} outperforms the source-channel separation scheme when one uses JPEG or JPEG2000 as the source code and LDPC code as the channel code for additive white Gaussian channel and fading channel with Rayleigh distribution. Furthermore, the authors show that while the classical approach suffers from the ``cliff effect,'' the proposed DL-based method's performance degrades gracefully as SNR decreases.

In this work, we have made two main contributions. First, by considering a Gaussian point-to-point wireless channel, we employ a variational auto-encoder (VAE) model instead of the standard AE (that has been thoroughly investigated in \cite{DeepJSCC_Gunduz_TCCN19}). In particular, our findings demonstrate that with the help of a VAE, we can obtain better results when the channel SNR degrades significantly.

Moreover, in this work, instead of the peak signal-to-noise ratio (PSNR), we use the structural similarity index (SSIM) \cite{ssim} as a perception-based metric to be maximized in the neural network. We have chosen SSIM measure because, in many real-world applications, the image transmitted through the channel would finally be used by humans. As discussed in \cite{loss_for_nn}, the PSNR would not be a proper metric in cases where human use is the goal. Additionally, SSIM would be a feasible choice because it can be calculated in a reasonable time, and its derivatives can be easily computed in the gradient descent algorithm \cite{loss_for_nn}. Also, it is demonstrated in \cite{generate_with_perceptual} that using perceptual-based metrics to reconstruct images can lead to better results in content-related tasks like classification rather than metrics like PSNR.

The rest of the paper is organized as follows.
Section~\ref{sec:Prob_Form} introduces the notation used in the paper and the problem formulation. We present our proposed method in Section~\ref{sec:Prop_Method}. The experimental results of the proposed method are presented in Section~\ref{sec:Exp_Results}. Finally, Section~\ref{sec:conclusion} concludes our paper.

\begin{figure}
	\centering
	\includegraphics[width=0.45\textwidth]{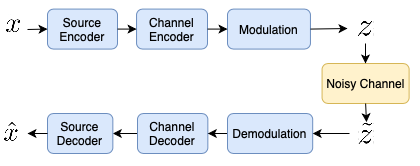}
	\caption{Classical scheme in digital data transmission through a noisy channel.}
	\captionsetup{justification=centering}
	\label{fig:classical_transmission}
\end{figure}

\section{Problem Formulation}\label{sec:Prob_Form}
\subsection{Notation}
We let $\mbb{R}$ and $\mbb{C}$ denote the set of real and complex numbers, respectively. Also, the multivariate Gaussian noise with mean vector $\mu$ and covariance matrix $m$ is denoted by $\mc{N}(\mu,m)$. We let $n$ and $k$ denote the dimension of input (\ie, size of the images) and the dimension of the transmitted data over the channel, respectively.

\subsection{Problem Statement}
In this paper, we address the problem of image transmission through a point-to-point communication channel. The images are represented by 2-dimensional arrays of pixel values. Each pixel consists of three integer values between 0 and 255, determining the color of the pixel in RGB mode. The first step of transmission in the \emph{classical} approach is to compress the image and reduce its size, using algorithms like JPEG or JPEG2000 known as \emph{source encoder}. The result is then fed to a \emph{channel encoder} (\ie, Turbo codes, LDPC codes) to protect the data from corruption from channel noise. The last step in this scheme is modulation. The stream of bits is converted to real or complex-valued signal parameters, with methods such as BPSK and 16-QAM. \autoref{fig:classical_transmission} illustrates all stages in such a classical approach.

To be more precise, we denote the input image by the vector $x\in\mbb{R}^n$ and the channel input by the vector $z\in\mbb{C}^k$, where we usually have $n>k$. We assume that the input to the channel satisfies the average power constraint defined by
\begin{equation}\label{eq:power_constrnt}
    \frac{1}{k}\mbb{E}[z^* z] \leq P.
\end{equation}
Then, the vector $z$ is transmitted through a noisy channel. This can lead to a change in the values of the vector $z$. So we denote the received vector in the decoder by $\tilde{z}\in\mbb{C}^k$. The received values are then demodulated and decoded by channel and source decoder to obtain an estimation of the original image denoted as $\hat{x}\in\mbb{R}^n$ (see Figure~\ref{fig:classical_transmission}).

By optimizing different modules of Figure~\ref{fig:classical_transmission}, the classical approach can achieve near-optimal performance. However, this requires an accurate design of each part for the specific operational parameters. In contrast, as mentioned in the Introduction, if one employs DL techniques to design an end-to-end communication system, the whole system optimization (including the joint source-channel coding) can be done together. Not only the DL based method is far easier than the analytical approach, but it has some other appealing characteristics. For example, it seems that the cliff effect has been handled better in this new approach \cite{DeepJSCC_Gunduz_TCCN19}.

In the next section, we will propose a VAE based method that models an end-to-end image transmission system over a wireless channel.

\section{The Proposed Method}\label{sec:Prop_Method}
The problem of transmitting data over noisy Gaussian channels has been investigated thoroughly after Shannon's paper \cite{shannon1948}. The established deep learning methods \cite{DeepJSCC_Gunduz_TCCN19} have provided us with good results with the image transmission problem. However, the methods proposed in \cite{DeepJSCC_Gunduz_TCCN19} lacks the required denoising ability to transmit images in lower SNR regimes. Therefore, further improvements and fine-tunings are needed to be made in order to have better results.

The main part of our proposed method involves the use of variational auto-encoders (VAE) instead of regular auto-encoders (AE). In the following, we will discuss the motivations behind this method and a brief description of VAEs.

VAE was first developed as a generative deep learning method \cite{VAE_Kingma_ICLR14}, that was an extension to the already developed auto-encoders by then. Even though it could not beat the state-of-the-art Generative Adversarial Networks (GAN) \cite{goodfellow_GAN_2014}, it inspired new ways and methods to solve other problems. In our method, we utilize VAEs' denoising capabilities to denoise and decode the transmitted symbols over a noisy channel.

\begin{figure}
	\centering
	\includegraphics[width=0.45\textwidth]{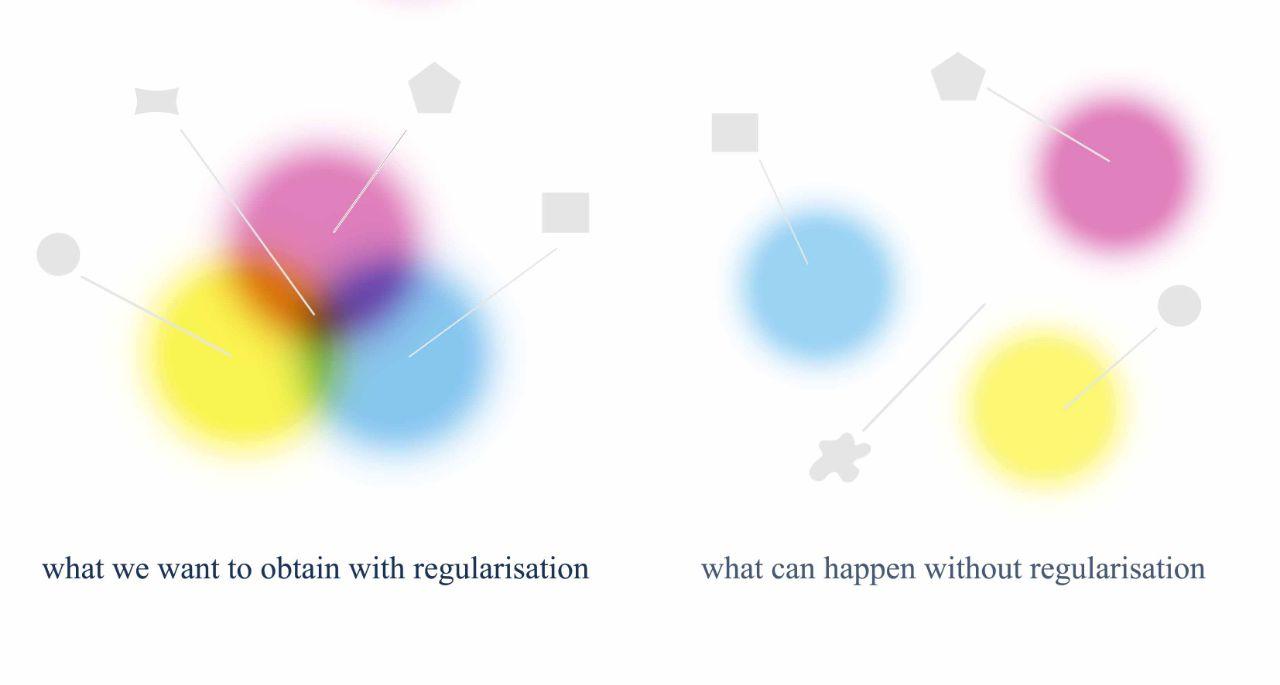}
	\caption{Regularized vs. unregularized latent space distribution.}
	\captionsetup{justification=centering}
	\label{fig:latent_space_reg}
\end{figure}

The main idea behind VAE is to come up with a way to force the auto-encoder's latent space to have a specific distribution. By choosing a smart loss function, we can impose the latent space distribution to be a standard Gaussian distribution. We will shortly discuss the reason behind selecting this probability distribution and how it can help us. In practice, we can achieve this goal by adding a KL-divergence term to the existing mean square error (MSE) loss as follows (see \cite{VAE_Kingma_ICLR14})
\[
    L = \|x-\hat{x}\|^2 + D_{KL}(h\| \mathcal{N}(0,I)),
\]
where $h$ is the latent variable vector. 
This will help the network to regularize the latent space distribution. Now one of the most important questions is, how regularizing the latent space can help us in denoising the auto-encoder? Imagine the simplified case, where our latent space has three clusters, as depicted in Figure~\ref{fig:latent_space_reg}. As one can see in the case of irregular latent space, data points that lie on a line between two distribution clusters but not quite close to each one, are meaningless and undecodable. However, in the case of regular latent space, these data points are more meaningful, and therefore, intuitively, we expect our network to be able to decode the data. Every image in our dataset can be roughly mapped to one of these clusters and can be decoded perfectly by an auto-encoder, but when we add the Gaussian noise channel to the problem, the auto-encoder seems to be not working quite well, especially in the case of the lower signal-to-noise ratio. In this case, all data points in the latent space experience a random relocation with respect to their first location. However, VAE's regularized latent space helps the decoder to be still able to decode the data point, and thus the results will be improved.

In order to be able to have more control over the extent of regularization in latent space, we can add a coefficient $\beta$ as a hyper-parameter to our loss function as follows
\[
    L = \|x-\hat{x}\|^2 + \beta D_{KL}(h \| \mc{N}(0,I)).
\]

\begin{figure}
	\centering
	\includegraphics[width=0.45\textwidth,height=4.8cm]{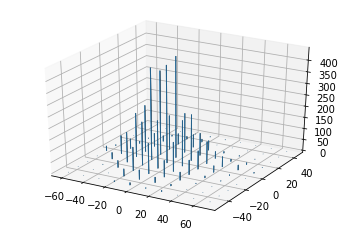}
	\caption{The histogram of an auto-encoder latent space, drawn with a two-component PCA.}
	\captionsetup{justification=centering}
	\label{fig:latent_dist}
\end{figure}

The main motive behind choosing the Gaussian as our VAE's latent space distribution stems from two facts. First, we know the optimal input distribution to an additive white Gaussian channel should be Gaussian. Second, the information provided in Figure~\ref{fig:latent_dist} also confirms that an AE provides a Gaussian distribution to the channel input. After analyzing the results of the AE and plotting the distribution of the reduced sized latent variables using the principal component analysis (PCA) method, we realized that the joint distribution of two leading PCA components of latent variables is approximately a (non-standard) Gaussian distribution. As we know, VAE forces the latent variable distribution to be a standard Gaussian distribution. This method works best if the auto-encoder distribution is a Gaussian distribution itself, and not some other distributions such as Gaussian mixture models. Using this information and the fact that VAE can help our model in denoising, the proposed method seems to be a natural practical solution for our problem.

In Section~\ref{sec:Exp_Results}, we propose a VAE architecture and test our proposed method on the STL-10 \cite{STL_10_dataset} and CIFAR-10 \cite{CIFAR_10_dataset} image datasets. We compare the results with a baseline auto-encoder architecture. The results will be tested with both PSNR and SSIM performance measures.

\subsection{Structural Similarity Index}
Structural similarity index (SSIM) \cite{ssim} is a perception-based metric to compare the quality of an image based on a ground truth concerning the human visual system (HVS). SSIM is a fully-referenced assessment method, meaning that the original image is completely needed in order to measure the quality of the reconstructed image. This method, unlike pixel-wise measures, \ie, MSE and mean absolute error (MAE), investigates the quality of the reconstructed image based on groups of pixels. The SSIM index for two signals $x$ and $y$ can be calculated as
\[
    \SSIM(x, y) \triangleq [l(x, y)]^\alpha \cdot [c(x, y)]^\beta \cdot [s(x, y)]^\gamma,
\]
where $l$, $c$ and $s$ are luminance, contrast and similarity comparison functions, respectively. If we assume $\mu_x$ and $\mu_y$ are the mean values and $\sigma_x$ and $\sigma_y$ are standard deviation and $\sigma_{xy}$ is the covariance of two signals, the we have
\begin{align*}
l(x, y) &\triangleq \frac{2\mu_x \mu_y + C_1}{{\mu_x}^2 {\mu_y}^2 + C_1},\\
c(x, y) &\triangleq \frac{2 \sigma_x \sigma_y + C_2}{{\sigma_x}^2 {\sigma_y}^2 + C_2},\\
s(x, y) &\triangleq \frac{\sigma_{xy} + C_3}{\sigma_x \sigma_y + C_3},
\end{align*}
where $C_1$, $C_2$ and $C_3$ are some constant numbers.

We also tried to improve the perceptual quality of the network's results by using SSIM to train the network. Inspired by \cite{loss_for_nn}, we use the following mixed loss function
\begin{equation}\label{eq:MixedLossFunc}
    \mathcal{L} \triangleq \alpha \mathcal{L}^{\SSIM} + (1-\alpha) \mathcal{L}^{L_1},
\end{equation}
where $\alpha\in[0,1]$ is a tuning parameter, and two cost functions $\mathcal{L}^{\SSIM}$ and $\mathcal{L}^{L_1}$ are defined as follows
\[
    \mathcal{L}^{\SSIM} \triangleq 1 - \SSIM(x, \hat{x}),
\]
\[
    \mathcal{L}^{L_1} \triangleq \frac{1}{n} \sum_{i=0}^{n}{|x_i - \hat{x}_i|}.
\]
As mentioned in \cite{loss_for_nn}, the choice of $\alpha$ has a significant impact on the quality of the results. The $\mc{L}^{L_1}$ loss, known as \emph{mean absolute error (MAE)}, is also added so to prevent the image colors from shifting.

We used the mixed loss function $\mathcal{L}$ proposed in \eqref{eq:MixedLossFunc} as both the loss function for auto-encoder and the reconstruction loss in variational auto-encoder.

\section{Experimental Results}\label{sec:Exp_Results}
\subsection{Model Architecture}
\begin{figure}
	\centering
	\includegraphics[width=0.45\textwidth]{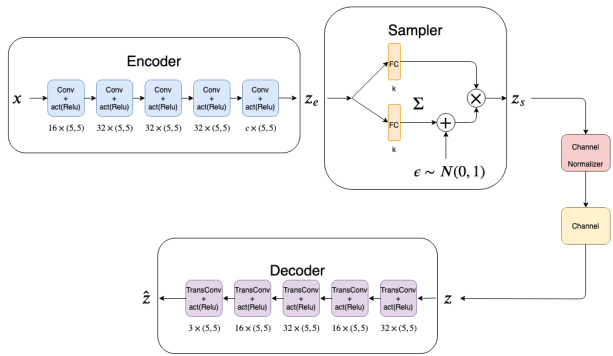}
	\caption{The architecture of the proposed VAE model.}
	\captionsetup{justification=centering}
	\label{fig:our_proposed_model}
\end{figure}

To demonstrate our proposed VAE method for the joint source-channel coding scheme, we use the deep neural network (NN) architecture depicted in Figure~\ref{fig:our_proposed_model}. In this scheme, we use an AWGN channel as the model for the channel noise, formulated as followed
\[
    \tilde{z}_i = z_i + \mc{N}(0, \sigma^2),\quad i \in [1:k],
\]
where index $i$ indicates the $i$th component of the vector $z$. The value of $\sigma$ determines the noise power and therefore the  channel SNR is given by
$\SNR = 10 \log\left(\frac{P}{\sigma^2}\right)$, where $P$ is the average power per symbol transmitted through the channel.

Our VAE model mainly consists of four parts: (1) encoder, (2) decoder, (3) channel normalizer, and (4) a sampler layer, as shown in Figure~\ref{fig:our_proposed_model}. The encoder's architecture consists of multiple layers of convolutional layer stacked on top of each other with parametric ReLu (PReLu) activation functions inter-connecting the layers. The goal of the encoder is to learn a good encoding of the data by reducing the size of the data and meanwhile learning important visual features. The decoder has the exactly mirrored architecture of the encoder, similar to all auto-encoder architectures.

Sampler layer consists of two parallel fully connected layers, which take as input the output of the last convolutional layer and produce as output, the mean vector $\mu$ and covariance matrix $\Sigma$. Followed by a sampler is a random generator that generates a random vector $\epsilon$ according to a standard Gaussian distribution $\mathcal{N}(0,I)$. Using the reparametrization trick introduced in \cite{VAE_Kingma_ICLR14}, we can generate the output of the sampler as follows $z_s = \mu + \Sigma \epsilon$.

Channel normalizer serves as a layer to enforce the power constraint on the AWGN channel. The encoder maps the $n$-dimensional input image $x$ to a $k$-dimensional vector of channel input samples $z$, which satisfies the average power constraint \eqref{eq:power_constrnt}. To satisfy this constraint, the output of the sampler is normalized according to
$z = \sqrt{kP}\frac{z_s}{\sqrt{z_s^* z_s}}$, where $z_s$ is the output of the sampler.

\subsection{Model Complexity}
As the baseline method, we developed an auto-encoder network that consists of five layers of convolution in the encoder and the same architecture mirrored in the decoder. Each convolutional layer consists of $D$ filters of size $W\times H$, so the number of weights for a convolutional layer would be $W\times H\times D$. Hence, considering all layers, the total number of parameters for the baseline auto-encoder approximately equals to $10^4$.

The proposed VAE method has the same architecture but with two extra fully-connected layers in the sampler. Fully-connected layers add a considerable amount of complexity to the architecture due to the fact that the number of weights for these layers is of order $O(N_1\times N_2)$ where $N_1$ and $N_2$ are the layer's input and output size, respectively. The total number of weights for the proposed VAE architecture would be close to $10^6$, which has increased by a factor of $100$. This would cause a significant reduction in training speed. In order to avoid this problem, a small trick can be used to still get the same result by increasing the training speed but sacrificing a small part of the network's capacity. Instead of using a fully-connected layer, we use another convolutional layer to acquire the same parameters in the sampler. This will help us achieve almost the same results in less time but would potentially decrease the final achievable accuracy of our model.

\subsection{Model Evaluation on CIFAR-10 with PSNR Metric}
Now that our model is ready, we evaluate the proposed method on the CIFAR-10 image datasets. We train the auto-encoder and VAE architectures on both datasets with varying channel conditions (SNR). The compression rate used to train models is $\frac{k}{n} = \frac{1}{6}$. During the training process, we use \emph{batch normalization} layers after each convolution layer. The training process continued for 1000 epochs. The initial learning rate is set to $10^{-3}$ and decayed to $10^{-4}$ after 350 epochs.

\begin{figure}
	\centering
	\includegraphics[width=0.45\textwidth,height=7cm]{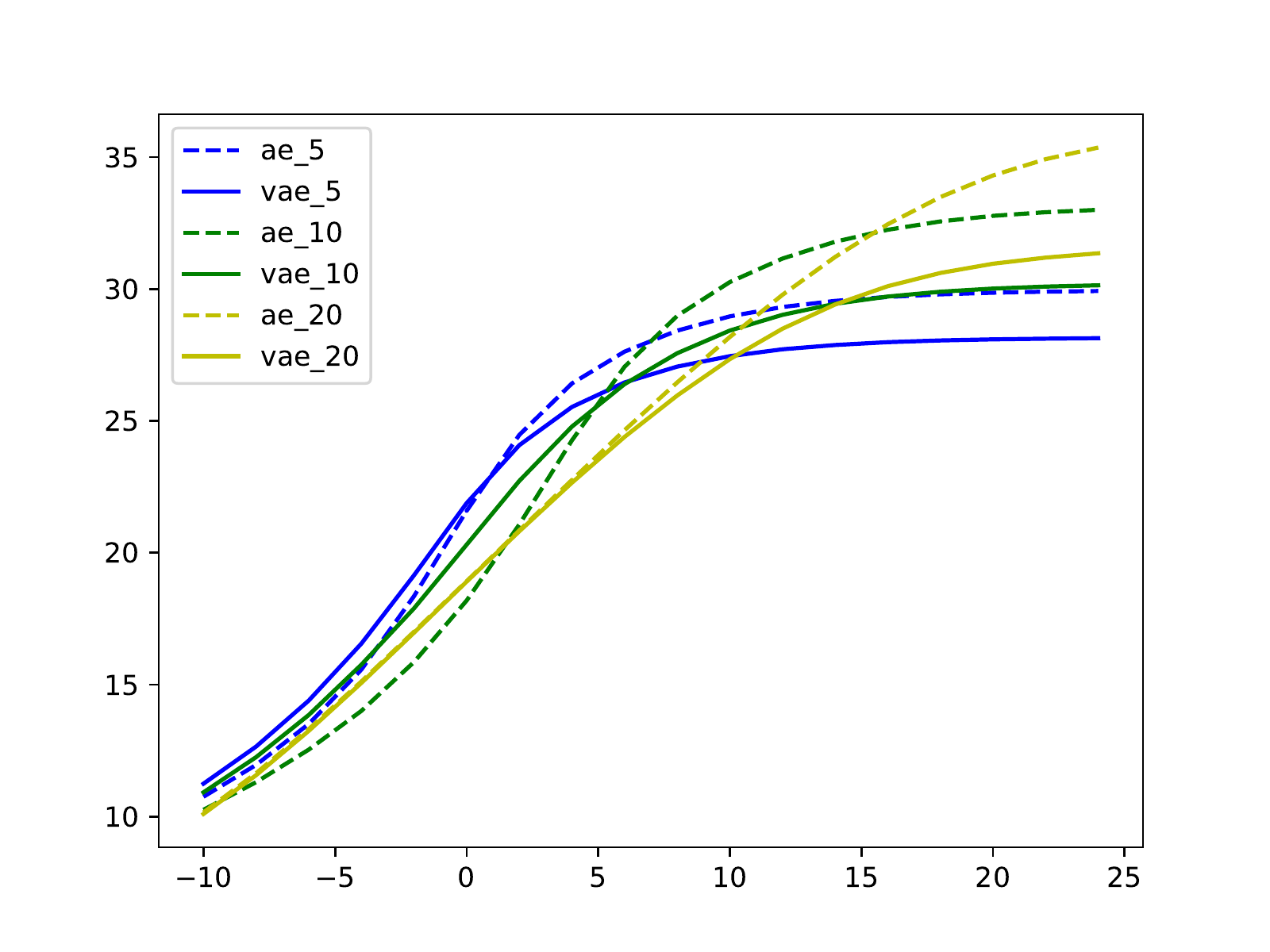}
	\caption{PSNR comparison of VAE vs. AE for different test SNRs.}
	\captionsetup{justification=centering}
	\label{fig:vae_vs_ae}
\end{figure}

\begin{figure}
	\centering
	\includegraphics[width=0.45\textwidth,height=7cm]{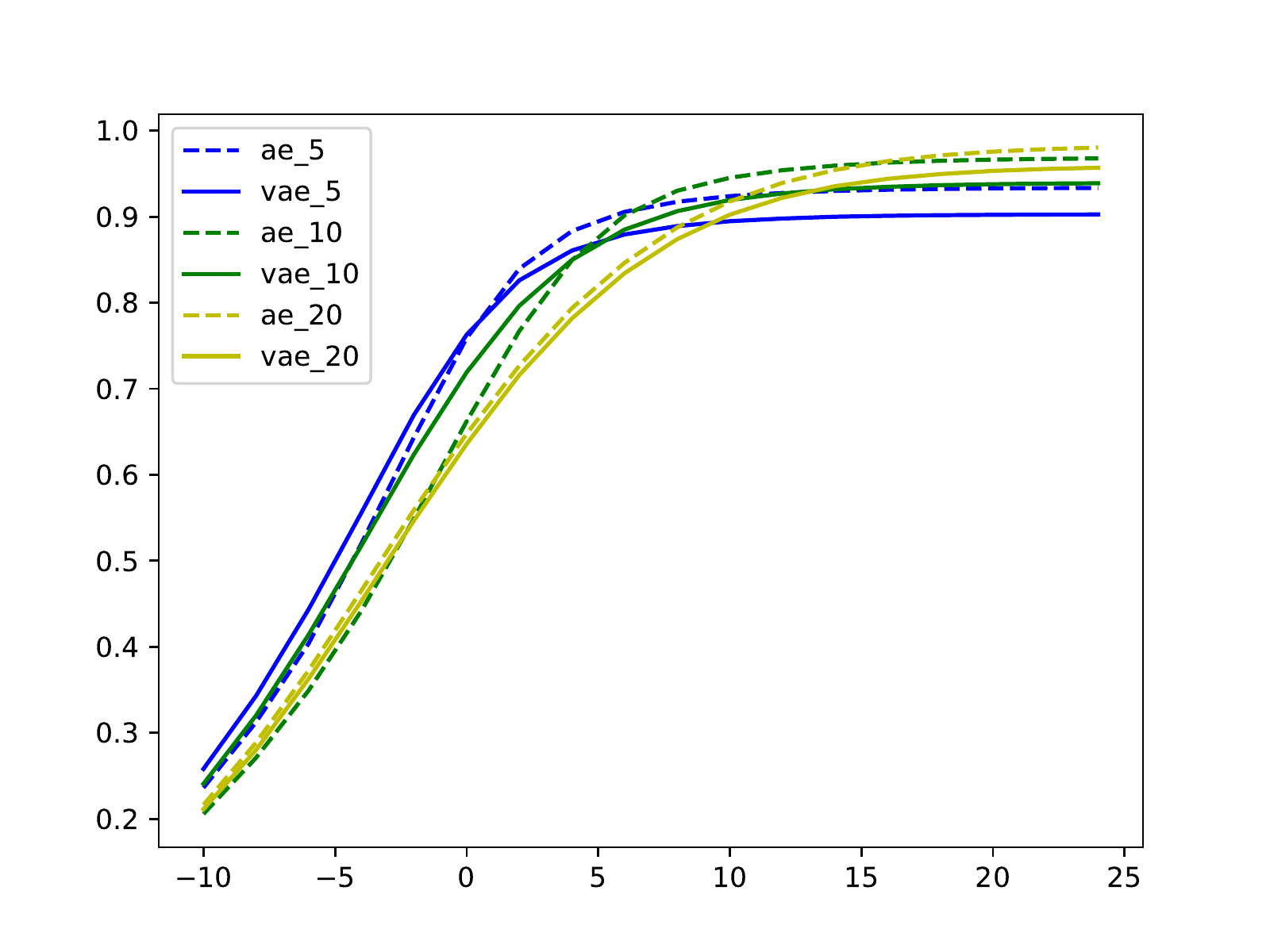}
	\caption{SSIM comparison of VAE vs. AE for different test SNRs.}
	\captionsetup{justification=centering}
	\label{fig:vae_vs_ae_ssim}
\end{figure}

\autoref{fig:vae_vs_ae} and \autoref{fig:vae_vs_ae_ssim} illustrate the performance of proposed VAE and AE models under different test SNRs. It can be observed that when the channel condition declines below the train SNR, the VAE model outperforms regular AE. VAE is a generative model that tries to learn the distribution of data. In contrast, AE learns a mapping from the input to output data. The process of learning the data distribution leads to a more generalization by the VAE, which explains its lower performance for the design SNR. However, when the channel degrades, the generalization that VAE has acquired help it outperforms the VE model.

\subsection{Model Evaluation on STL10 with SSIM Metric}
Due to the low quality of images in the CIFAR-10 dataset, investigation of the perceptual quality of the model on this data would be impossible. Therefore we trained the model on the STL10 dataset that has a better quality comparing to CIFAR10. We have used two different loss functions to train both auto-encoder and variational auto-encoder. The loss functions we used are MSE and the proposed L1-SSIM loss function with $\alpha$ set to $\frac{1}{2}$. This value is chosen intuitively, and we did not perform an investigation on the values of $\alpha$. All four models are trained on 100,000 images with 150 epochs with $\SNR = 10$dB and $\frac{k}{n} = \frac{1}{6}$.

\begin{figure*}
    \centering
    \captionsetup{width=.85\linewidth, justification=centering}
    \includegraphics[width = 0.9 \linewidth]{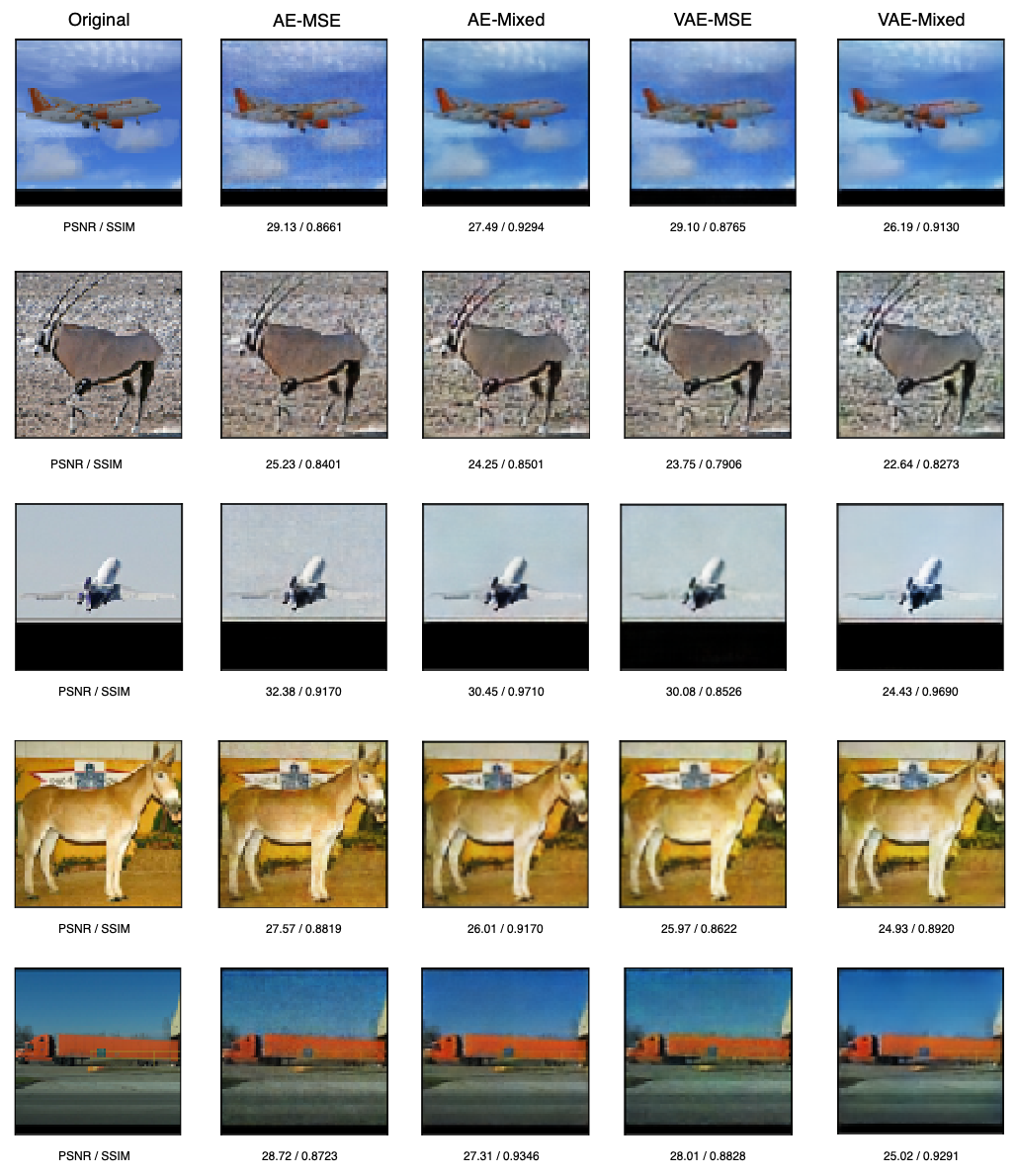}
    \caption{Performance of VAE and AE with MSE and proposed mixed loss function $\mc{L}$ defined in \eqref{eq:MixedLossFunc}. Each model is trained for 150 epochs with 100000 train images with both test and train channel quality of $\SNR=10$dB.}
    \label{fig:ssim_images}
\end{figure*}

\autoref{fig:ssim_images} illustrates five images generated with these four proposed models. As expected, images generated with models trained using mixed MAE-SSIM loss are sharper and more clear. Also, MSE loss, whose objective is to maximize PSNR, leads to more blurry images, especially in small details. Furthermore, models trained with mixed loss perform better under wide areas of a specific color due to its power to eliminate luminance noise in such areas.  We also found out that while using MSE as loss function, images generated by VAE have better quality. We suppose VAE due to its natural structure is more resilient to noise and has a smoother output. On the other hand, MSE keeps the colors more chromatic. It is necessary to mention that fine-tuning $\alpha$ with respect to the situation can have a significant impact on the quality of the outcome.

{}

\section{Conclusion}\label{sec:conclusion}
In this work, we have proposed using a variational auto-encoder (VAE) structure to model an end-to-end wireless communication system. Our findings have shown that the proposed VAE based scheme outperforms the scheme based on using a standard AE in the lower SNR regime. Moreover, we have observed that using SSIM measure instead of PSNR measure resulted in better visually reconstructed images at the receiver.

\section*{Acknowledgment}
The authors would like to thank Seyed Pooya Shariatpanahi for many useful feedback he kindly provided on this work.

\bibliography{ML4COM} 
\bibliographystyle{ieeetr}

\end{document}